\documentclass[a4paper,11pt]{article}

\pdfoutput=1

\usepackage{amsmath}
\usepackage{amssymb}
\usepackage{graphicx,color}
\usepackage{hyperref}
\usepackage[a4paper,left=0.99in, right=0.99in,top=1.1in, bottom=1.2in]{geometry}
\usepackage{cite}

\numberwithin{equation}{section}

\usepackage{array}
\usepackage{multirow}
\usepackage{setspace}
\usepackage{xspace}

\newcommand{\that}{{\hat t}}
\newcommand{\shat}{{\hat s}}
\newcommand{\uhat}{{\hat u}}
\newcommand{\tlt}{\bar{t}}
\newcommand{\tls}{\bar{s}}
\newcommand{\tlu}{\bar{u}}

\newcommand{\Ei}{\text{Ei}}
\def\be{\begin{equation}}
\def\ee{\end{equation}}
\def\bea{\begin{eqnarray}}
\def\eea{\end{eqnarray}}

\title{ 
    \vskip 2cm
    Forward di-jet production in p+Pb collisions in the \\ 
    small-$x$ improved TMD factorization framework
}

\author{
A. van Hameren$^2$, P. Kotko$^1$, K. Kutak$^2$, C. Marquet$^3$, 
E. Petreska$^{3,4}$ and  S.  Sapeta$^{5,2}$\\\\
$^1$ {\small\it Department of Physics, Penn State University}\\ {\small\it University Park, 16803 PA, USA}\\\\
$^2$ {\small\it The H.\ Niewodnicza\'nski Institute of Nuclear Physics PAN}\\ {\small\it Radzikowskiego 152, 31-342 Krak\'ow, Poland}\\\\
$^3$ {\small\it Centre de Physique Th\'eorique, \'Ecole Polytechnique,}\\
{\small\it CNRS, Universit\'e Paris-Saclay, 91128 Palaiseau, France}\\\\
$^4$ {\small\it Departamento de F\'isica de Part\'iculas and IGFAE,}\\
{\small\it Universidade de Santiago de Compostela, 15782 Santiago de Compostela, Spain}\\\\
$^5$ {\small\it Theoretical Physics Department, CERN, Geneva, Switzerland  }\\
}

\begin{document}

\date{}

\maketitle

%--------------- preprint numbers ---------------------
\vspace{-40em}
\begin{flushright}
  CERN-TH-2016-159 \\
  CPHT-RR036.072016 \\
  IFJPAN-IV-2016-18
\end{flushright}
\vspace{35em}
%-----------------------------------------------------------------

\abstract{

We study the production of forward di-jets in proton-lead and proton-proton collisions at the Large Hadron Collider.
Such configurations, with both jets produced in the forward direction, impose a dilute-dense asymmetry which 
allows to probe the gluon density of the lead or proton target at small longitudinal momentum fractions.
Even though the jet momenta are always much bigger than the saturation scale of the target, $Q_s$, the transverse momentum
imbalance of the di-jet system may be either also much larger than $Q_s$, or of the order $Q_s$, implying that the small-$x$ QCD
dynamics involved is either linear or non-linear, respectively. The small-$x$ improved TMD factorization framework deals with
both situation in the same formalism. In the latter case, which corresponds to nearly back-to-back jets, we find that saturation
effects induce a significant suppression of the forward di-jet azimuthal correlations in proton-lead versus proton-proton collisions.

}

%\tableofcontents
\clearpage

%%%%%%%%%%%%%%%%%%%%%%%%%%%%%%%%%%%%%%%%%%%%%%%%%%%%%%%%%%%%%%%%%%%%%%%%%%%% 
\section{Introduction}

Measurements of forward particle production in high-energy hadronic collisions provide unique
opportunities to study the QCD dynamics of the non-linear parton saturation regime~\cite{Gribov:1984tu}.
Such processes, in which, for kinematical reasons, high-momentum partons from one of the colliding hadrons mainly scatter with small-momentum partons from the other, are called dilute-dense collisions. Indeed, the density of the large-$x$ partons in the projectile hadron is small, while the density of the small-$x$ gluons in the target hadron is large, and the former, well understood in perturbative QCD, can be used to probe the dynamics of the latter. This is true already in proton-proton collisions, although using a target nucleus does enhance the dilute-dense asymmetry of such collisions.
 
RHIC measurements have provided some evidence for the presence of saturation effects
in the data, the most compelling of which is the successful description of forward di-hadron
production \cite{Albacete:2010pg,Stasto:2011ru,Lappi:2012nh}, using the most up-to-date
theoretical tools available at the time in the Color Glass Condensate (CGC) framework \cite{Gelis:2010nm,Albacete:2014fwa}. In particular, this approach predicted the suppression of azimuthal correlations in d+Au collisions compared to p+p collisions \cite{Marquet:2007vb}, which was observed later experimentally \cite{Adare:2011sc,Braidot:2010zh}.

The CGC effective theory provides a tool to compute observables when non-linear QCD dynamics must be taken into account. It effectively describes, in terms of strong classical fields, the dense parton content of a hadronic/nuclear wave function, at small longitudinal momentum fraction $x$. The separation between the linear and non-linear regimes is characterized by a momentum scale $Q_s(x)$, called the saturation scale, which increases as $x$ decreases, and roughly scales as $A^{1/6}$. The CGC description of dilute-dense collisions from first principles is valid provided $Q_s\gg\Lambda_{\mathrm{QCD}}$, therefore it should work better with higher energies, as they open up the phase space towards lower values of $x$. In order to verify that this is the case, the CGC predictions must be extended from RHIC kinematics to the Large Hadron Collider (LHC), where the relevant observables involve high-$p_t$ jets, as opposed to individual hadrons with $p_t$ of the order
of a few GeV at RHIC.

In this context, we shall consider forward di-jet production in proton-lead versus proton-proton collisions. In that case, it was shown in \cite{Kotko:2015ura} that the full complexity of the CGC machinery is not needed. Indeed, for the di-hadron process at RHIC energies, no particular ordering of the momentum scales involved is assumed in CGC calculations, while at the LHC one can take advantage of the presence of final-state partons with transverse momenta much larger than the saturation scale to obtain simplifications. On the flip side, different complications - left for future studies - are expected to arise due to QCD dynamics relevant at large transverse momenta and not part of the CGC framework, such as Sudakov logarithms \cite{Mueller:2012uf,Mueller:2013wwa,vanHameren:2014ala,Kutak:2014wga} or coherence in the QCD evolution of the gluon density \cite{Ciafaloni:1987ur,Catani:1989sg,Catani:1989yc}.

There are three distinct momentum scales in the forward di-jet process. The typical jet transverse momentum $P_t$ is always one of the hardest scales, and it is much bigger than the saturation scale $Q_s$, which is always one of the softest scales. The third momentum scale is the total transverse momentum of jet pair $k_t$, which also corresponds to the transverse momentum of the small-$x$ gluons involved in the hard scattering. Depending on where $k_t$ sits with respect to $P_t$ and $Q_s$, the full CGC formulation simplifies either to the \emph{high energy factorization} (HEF) framework \cite{Catani:1990eg,Deak:2009xt} or to the (small-$x$ limit of the) \emph{transverse momentum dependent} (TMD) factorization framework \cite{Bomhof:2006dp}.

The HEF framework is recovered from the CGC when $Q_s\ll k_t\sim P_t$ \cite{Kotko:2015ura}. In that case, non-linear effects are absent, and the description of forward di-jets involves off-shell hard matrix elements, along with a single TMD gluon distribution for the small-$x$ target (also called unintegrated gluon distribution in the literature). The TMD framework is recovered from the CGC when $k_t\sim Q_s\ll P_t$ \cite{Dominguez:2010xd,Dominguez:2011wm,MPR}. In that case, some non-linear effects do survive, and the description of forward di-jets involves several TMD gluon distributions, each associated to a sub-set of the hard matrix elements, but those are on-shell. In Ref.\cite{Kotko:2015ura}, we proposed an interpolating formula between those two limits, applicable for $P_t\gg Q_s$ regardless of the magnitude of $k_t$, which is more amenable to phenomenological implementations than the CGC expression which, as we pointed out, also contains both the HEF and TMD limits. Not unexpectedly, it involves several unintegrated gluons distributions each associated to a sub-set of off-shell matrix elements. The goal of this paper is to provide a numerical implementation of that new formulation, dubbed improved TMD (ITMD) factorization.

The off-shell matrix elements needed to compute the forward di-jet process have all been calculated in \cite{Kotko:2015ura},
but evaluating all the necessary gluon TMDs is not straightforward. Very recently, they have been obtained from a numerical simulation of the non-linear QCD evolution in the leading $\ln(1/x)$ approximation \cite{MPR}, that is from the Jalilian-Marian-Iancu-McLerran-Weigert-Leonidov-Kovner (JIMWLK) \cite{JalilianMarian:1997jx,JalilianMarian:1997dw,Iancu:2000hn,Ferreiro:2001qy,Weigert:2000gi} equation. However, further work is required before those TMDs can be incorporated into a cross section calculation. Therefore, in the present work, we shall stick to a mean-field type approach in which all the gluon distributions needed can be related to each other, and obtained from the simpler Balitsky-Kovchegov (BK) equation \cite{Balitsky:1995ub,Kovchegov:1999yj}. A detailed comparative study using solutions of the different extensions of the original BK equation is left for future work. The version that we shall use in this work is known as the KS gluon distribution \cite{Kutak:2012rf}. It incorporates the running of the QCD coupling, non-singular pieces (at low $x$) of the DGLAP splitting function, a sea-quark contribution, and resums dominant corrections from higher orders via a kinematic constraint \cite{Kwiecinski:1997ee, Kutak:2003bd}.

By comparing the forward di-jet production cross sections in proton-lead and proton-proton collisions, we can clearly see the onset of parton saturation effects, as we go from a kinematical regime in which $k_t\sim P_t$ towards one where $k_t\sim Q_s$, and we obtain a good estimation of the size of those effects where they are the biggest, which is for nearly back-to-back jets. We note that probing non-linear effects of similar strength with single-inclusive observables requires to make the only transverse momentum involved in those processes of the order of the saturation scale, which may not be easy experimentally. With di-jets, assuming $P_t\sim 20$ GeV and $k_t\sim Q_s \sim 2$ GeV, we can reach $R_{pPb}\sim 0.5$.

The paper is organized as follows. In section 2, we recall the essence and the ingredients of the ITMD factorization formula for forward di-jets in dilute-dense collisions. In section 3, we introduce the mean-field approximation that allows us to express the various gluon TMDs in terms of the solution of the BK equation. In section IV, we present numerical results for the proton and lead gluon TMDs obtained with the KS gluons, and compared them with analytical expressions obtained in the GBW model. In section V, we present our results for forward di-jet production in p+p and p+Pb collisions at the LHC, as well as nuclear modification factors $R_{\rm pPb}$. Finally, section VI is devoted to conclusions and outlook.

\section{The ITMD factorization formula for forward di-jets in dilute-dense collisions}

We consider the process of inclusive forward di-jet production in hadronic collisions
\begin{equation}
  p (p_p) + A (p_A) \to j_1 (p_1) + j_2 (p_2)+ X\ ,
\end{equation}
where the four-momenta of the projectile and the target are massless and purely longitudinal.
The longitudinal momentum fractions of the incoming parton from the projectile, $x_1$, and the gluon from the target, $x_2$, can be expressed in
terms of the rapidities $(y_1,y_2)$ and transverse momenta $(p_{t1},p_{t2})$ of the produced jets as
\be
x_1  = \frac{p_1^+ + p_2^+}{p_p^+}   = \frac{1}{\sqrt{s}} \left(|p_{1t}|
e^{y_1}+|p_{2t}| e^{y_2}\right)\ , \quad
x_2  = \frac{p_1^- + p_2^-}{p_A^-}   = \frac{1}{\sqrt{s}} \left(|p_{1t}| e^{-y_1}+|p_{2t}| e^{-y_2}\right)\ .
\ee
By looking at jets produced in the forward direction, we effectively select those fractions to be $x_1 \sim 1$ and $x_2 \ll 1$.
Since the target A is probed at low $x_2$, the dominant contributions come from the subprocesses in which the incoming parton on the target side is a gluon
\begin{equation}
  qg  \to  qg\ ,
  \qquad \qquad 
  gg  \to  q\bar q\ ,
  \qquad \qquad 
  gg  \to  gg\ .
\label{eq:3channels}
\end{equation}

Moreover, the large-$x$ partons of the dilute projectile are described in terms of the usual parton distribution functions of collinear factorization $f_{a/p}(x_1)$ while the small-$x$ gluons of the dense target are described by TMD distributions $\Phi_{g/A}(x_2,k_t)$ . Indeed, the momentum of the incoming gluon from the target is not only longitudinal but also has a non-zero transverse component of magnitude
\be
k_t = |p_{1t}+p_{2t}|
\label{eq:ktglue}
\ee
which leads to imbalance of transverse momentum of the produced jets: $k_t^2 =|p_{1t}|^2 + |p_{2t}|^2 + 2|p_{1t}||p_{2t}| \cos\Delta\phi$.
The validity domain of ITMD factorization is
\be
Q_s(x_2)\ll P_t
\ee
where $P_t$ is the hard scale of the process, related to the individual jet momenta $P_t \sim |p_{1t}|,|p_{2t}|$. By contrast, the value of $k_t$ can be arbitrary.

The ITMD factorization formula reads \cite{Kotko:2015ura}
\begin{equation}
\frac{d\sigma^{pA\rightarrow {\rm dijets}+X}}{d^{2}P_{t}d^{2}k_{t}dy_{1}dy_{2}}=\frac{\alpha_{s}^{2}}{(x_1 x_2 s)^{2}}
\sum_{a,c,d} \frac{x_1 f_{a/p}(x_1)}{1+\delta_{cd}}\sum_{i=1}^{2}K_{ag^*\to cd}^{(i)}(P_t,k_t)\Phi_{ag\rightarrow cd}^{(i)}(x_2,k_t)\ .
\label{eq:itmd}
\end{equation}
It involves several gluon TMDs $\Phi_{ag\rightarrow cd}^{(i)}$ (2 per channel), with different operator definitions, that are accompanied by different hard factors $K_{ag^*\to cd}^{(i)}$. Those where computed in \cite{Kotko:2015ura} using either Feynman diagram techniques, or color-ordered amplitude methods, and they are given in Table~\ref{tab:Khardfactors} in terms of the Mandelstam variables of the $2\to 2$ parton level process. They encompass the improvement over the TMD factorization formula derived in Ref.~\cite{Dominguez:2011wm} where the matrix elements were on-shell and a function of $P_t$ only. The gluon TMDs are normalized such that
\be
\int d^2k_t\ \Phi_{ag\rightarrow cd}^{(i)}(x_2,k_t)= x_2 f_{g/A}(x_2)\ ,
\ee
and their precise operator definitions can be found in \cite{Kotko:2015ura}.

\renewcommand{\arraystretch}{2.}
\begin{table}
\begin{doublespace}
\begin{centering}
\begin{tabular}{c|c|c}
\hline  
\Large $i$ & \Large 1 &  \Large 2  \\
\hline 
$\displaystyle K_{gg^*\to gg}^{(i)}$ & $ \displaystyle
2\,\frac{\left(\overline{s}^{4}+\overline{t}^{4}+\overline{u}^{4}\right)\left(\overline{u}\hat{u}+\overline{t}\hat{t}\right)}{\tlt\that\tlu\uhat\tls\shat}$ & $\displaystyle -\,\frac{\left(\overline{s}^{4}+\overline{t}^{4}+\overline{u}^{4}\right)\left(\overline{u}\hat{u}+\overline{t}\hat{t}-\overline{s}\hat{s}\right)}{\tlt\that\tlu\uhat\tls\shat}$
\\
\hline 
$\displaystyle K_{gg^*\to q\overline{q}}^{(i)}$ & $\displaystyle \frac{1}{2N_{c}}\,\frac{\left(\overline{t}^{2}+\overline{u}^{2}\right)\left(\overline{u}\hat{u}+\overline{t}\hat{t}\right)}{\overline{s}\hat{s}\hat{t}\hat{u}}$ & $\displaystyle \frac{1}{2N_{c}^3}\,\frac{\left(\overline{t}^{2}+\overline{u}^{2}\right)\left(\overline{u}\hat{u}+\overline{t}\hat{t}-\overline{s}\hat{s}\right)}{\overline{s}\hat{s}\hat{t}\hat{u}}$
\\
\hline 
$\displaystyle K_{qg^*\to qg}^{(i)}$ & $\displaystyle -\frac{\overline{u}\left(\overline{s}^{2}+\overline{u}^{2}\right)}{2\overline{t}\hat{t}\hat{s}}$ & $\displaystyle -\,\frac{\overline{s}\left(\overline{s}^{2}+\overline{u}^{2}\right)}{2\overline{t}\hat{t}\hat{u}}$
\\
\hline 
\end{tabular}
\par\end{centering}
\end{doublespace}
\caption{The hard factors accompanying the gluon TMDs $\Phi_{ag\rightarrow cd}^{\left(i\right)}$ in the large-$N_c$ limit. The finite $N_c$ expressions can be found in \cite{Kotko:2015ura}.}
\label{tab:Khardfactors}
\end{table}

As emphasized in the introduction, formula \ref{eq:itmd} coincides with CGC expressions in two important limits. They both reduce to the TMD factorization formula when $Q_s\sim k_t\ll P_t$ and to the HEF formula when $Q_s\ll k_t\sim P_t$:
\begin{itemize}
\item The TMD factorization formula with $k_t$ dependent gluon distributions and on-shell matrix elements is simply obtained form \ref{eq:itmd} after simplifying
$K_{ag^*\to cd}^{(i)}(P_t,k_t)$ into $K_{ag^*\to cd}^{(i)}(P_t,0)\equiv K_{ag\to cd}^{(i)}(P_t)$:
\begin{equation}
\frac{d\sigma^{pA\rightarrow {\rm dijets}+X}}{d^{2}P_{t}d^{2}k_{t}dy_{1}dy_{2}}=\frac{\alpha_{s}^{2}}{(x_1 x_2 s)^{2}}
\sum_{a,c,d} \frac{x_1 f_{a/p}(x_1)}{1+\delta_{cd}}\sum_{i=1}^{2}K_{ag\to cd}^{(i)}(P_t)\Phi_{ag\rightarrow cd}^{(i)}(x_2,k_t)\ .
\end{equation} 
The derivation of this expression from the CGC framework was done in \cite{Dominguez:2011wm} the in large-$N_c$ limit, and in \cite{MPR} for the finite $N_c$ case. However the TMD approach had been previously extensively studied in the literature \cite{Bomhof:2006dp,Boer:1999si,Belitsky:2002sm,Boer:2003cm,Collins:2007nk,Vogelsang:2007jk,Rogers:2010dm,Xiao:2010sp}, and in a broader context than small-x physics. 
\item Obtaining the HEF formula with a single gluon TMD and off-shell matrix elements from Eq.~\ref{eq:itmd} relies on the fact that up to power corrections, all the gluon TMDs coincide in the large $k_t$ limit:
\be
\Phi_{ag\rightarrow cd}^{(i)}(x_2,k_t)\to \Phi_{g/A}(x_2,k_t)+{\cal O}(1/k_t^2)\ .
\label{eq:highpt}
\ee
Then, denoting
\be
g_s^4\sum_{i=1}^{2}K_{ag^*\to cd}^{(i)}(P_t,k_t)=|\overline{{\cal M}_{ag^*\to cd}}|^2\, 
\ee
the HEF formula is
\begin{equation}
\frac{d\sigma^{pA\rightarrow {\rm dijets}+X}}{d^{2}P_{t}d^{2}k_{t}dy_{1}dy_{2}}=\
\frac{1}{16\pi^2 (x_1x_2 s)^2} \sum_{a,c,d} \frac{x_1 f_{a/p}(x_1)}{1+\delta_{cd}}
  |\overline{{\cal M}_{ag^*\to cd}}|^2  \Phi_{g/A}(x_2,k_t)\ .
  \label{eq:hef-formula}
\end{equation}
This expression is also obtained from the CGC framework in the dilute target limit \cite{Kotko:2015ura},
and has also been extensively studied in the literature \cite{Deak:2009xt,Kutak:2012rf,vanHameren:2014lna,vanHameren:2014ala,vanHameren:2013fla}
(where the gluon TMD is denoted ${\cal F}_{g/A}=\pi \Phi_{g/A}$ due to a different normalization convention).
\end{itemize}

We would like to point out that the ITMD factorization formula \ref{eq:itmd} was build in order to contain both the HEF and the TMD expressions as its limiting cases, and as such should be considered no more than an interpolating formula. We note however, that if one would be able to directly derive a factorization formula valid for $Q_s\ll P_t$ regardless of the value of $k_t$, any additional term compared to \ref{eq:itmd} should vanish in both limits $Q_s\sim k_t\ll P_t$ and $Q_s\ll k_t\sim P_t$.

\section{The gluon TMDs in the Gaussian approximation}

The goal of this paper is to provide a numerical implementation of the ITMD factorization formula, which first requires to evaluate all the gluon TMDs that enter
Eq.~\ref{eq:itmd}. Let us start with the simplest of them, $\Phi_{qg\rightarrow qg}^{(1)}$, also called the dipole gluon distribution and often denoted $x_2G^{(2)}$. In the small-$x_2$ limit, it can be related to the Fourier transform of the fundamental dipole amplitude $N_F(x_2,{\bf{r}})$ where $\bold{r}$ denote the transverse size of the dipole \cite{Dominguez:2011wm,MPR}:
\be
\Phi_{qg\rightarrow qg}^{(1)}(x_2,k_t)= \frac{N_c}{\alpha_s \pi (2\pi)^3}\int d^2b\int d^2 {\bf{r}}\
e^{-i k_t \cdot{\bf r}}\nabla^2_{{\bf{r}}}\ N_F(x_2, {\bf{r}}) \equiv x_2 G^{(2)}(x_2, k_t) \ .
\label{eq:dipgluontmd}
\ee
The amplitude $N_F$ is defined through the CGC expectation value of the $S$-matrix, $S_F$, of a quark-antiquark dipole scattering off the dense target: $N_F(x,{\bf{r}})= 1 - S_F(x, {\bf{r}})$ with $S_F(x, {\bf{r}})=\left< \text{Tr} \left [ U({\bold{r}})U^\dagger({\bold{0}})\right]\right>_x/N_c$ in terms of fundamental Wilson lines. The dipole gluon distribution can then be written in a compact form as:
\be 
x_2 G^{(2)}(x_2, k_t) = \frac{N_c\ k_t^2\ S_\perp}{2\pi^2 \alpha_s} F(x_2,k_t) \ ,
\label{eq:dipolegluon}
\ee
where $F(x_2,k_t)$ is a Fourier transform of the fundamental dipole
\be 
F(x_2,k_t) = \int \frac{d^2{\bf{r}}}{(2\pi)^2} e^{-i{k_t}\cdot{\bf r}} S_F(x_2, {\bf{r}}) ,
\ee
and with $S_\perp$ denoting the transverse area of the target.

In full generality, none of the other gluon TMDs can be obtained in such a straightforward manner. For instance, the Weizs{\"a}cker-Williams (WW) gluon distribution, denoted $x_2G^{(2)}$, should be obtained in the small-$x_2$ limit from the quadrupole operator $\left< \text{Tr} \left [ A({\bold{x}})A({\bold{y}})\right]\right>_{x_2}$ where $A({\bold{x}})=U^\dagger({\bold{x}})\partial_{\bf x}U({\bold{x}})$, and in general is not related to $F(x_2,k_t)$. Therefore, in order to simplify the evaluation of all the gluon TMDs which we need, we will resort to a mean-field type approximation.

We shall utilize the so-called Gaussian approximation of the CGC
\cite{Fujii:2006ab,Marquet:2007vb,Kovchegov:2008mk,Marquet:2010cf,Dumitru:2011vk,Iancu:2011nj,Alvioli:2012ba}. The essence of this approximation is to assume that all the color charge correlations in the target stay Gaussian throughout the evolution: $\left< \rho({\bold{x}})\rho({\bold{y}})\right>_{x}\!\propto\!\mu^2(x,{\bold{x}}-{\bold{y}})$. In addition, for simplicity, we shall work in the large-$N_c$ limit. This Gaussian approximation allows to write, among other things, the WW gluon distribution in terms of an adjoint dipole:
\be
x_2G^{(1)}(x_2,k_t)= \frac{C_F}{2 \alpha_s \pi^4}\int d^2b\int \frac{d^2{\bf r}}{{\bf r}^2}\
e^{-i{k_t}\cdot{\bf r}}\ \left[1 - S_A(x_2,{\bf r})\right] \ ,
\label{eq:WW_AdjointDipole}
\ee
where now, $S_A(x,{\bf{r}})$ is an $S$-matrix for the scattering of a gluon dipole involving adjoint Wilson lines.
The Gaussian approximation also allows to write $S_F=S_{BK}^{2C_F/C_A}$ and $S_A=S_{BK}^2$, where $C_F$ and $C_A$ are the Casimirs of the fundamental and adjoint representations of $SU(N_c)$, respectively, and with $S_{BK}$ denoting the solution of the BK equation. At large $N_c$, $S_A(x,{\bf{r}}) = \left[S_F(x,{\bf{r}})\right]^2$, and one can write:
\begin{eqnarray}
k_t^2 \nabla_{k_t}^2\ x_2G^{(1)}(x_2,k_t)&=&  \frac{C_F S_\perp}{2 \alpha_s \pi^4}\ k_t^2\int d^2{\bf r}\ e^{-i{k_t}\cdot{\bf r}}\ \left[S_F(x_2,{\bf r})\right]^2\\
&=& \frac{2 C_F S_\perp}{\alpha_s \pi^2}\ k_t^2\int d^2q_t\ F(x_2,q_t)F(x_2,k_t-q_t)\\
&=& 2k_t^2 \int \frac{d^2q_t}{q_t^2}\ x_2G^{(2)}(x_2,q_t)F(x_2,k_t-q_t).
\end{eqnarray}
Then the Laplacian can be inverted as:
\begin{equation}
x_2G^{(1)}(x_2,k_t)=\frac12 \int_{k_t^2}^{\infty}
dk_t^{'2}\ln\left(\frac{k_t'^{2}}{k_t^2}\right) \int \frac{d^2q_t}{q_t^2}\
x_2G^{(2)}(x_2,q_t)F(x_2,k_t'-q_t) \ .
\label{eq:WW(F)}
\end{equation}

In the large $N_c$ limit, the six gluon distributions $\Phi_{ag\rightarrow cd}^{(i)}$ reduce to \cite{Kotko:2015ura}:
\bea 
\Phi_{qg\to qg}^{(1)} = \mathcal{F}_{qg}^{(1)}~~~~~&,&~~~~~\Phi_{qg\to qg}^{(2)} \approx \mathcal{F}_{qg}^{(2)} \\
\Phi_{gg\to q\bar{q}}^{(1)} \approx \mathcal{F}_{gg}^{(1)}~~~~~&,&~~~~~\Phi_{gg\to q\bar{q}}^{(2)} \approx - N_c^2\mathcal{F}_{gg}^{(2)} 
\label{eq:phigg2}
\\
\Phi_{gg\to gg}^{(1)} \approx 
\frac{1}{2}\left(\mathcal{F}_{gg}^{(1)}+\mathcal{F}_{gg}^{(6)}\right)~~~~~&,&~~~~~
\Phi_{gg\to gg}^{(2)} \approx \mathcal{F}_{gg}^{(2)}+\mathcal{F}_{gg}^{(6)}
\eea
Therefore, we need an input of five gluon TMDs in our numerical calculations, the dipole gluon distribution, and four others: $\mathcal{F}_{qg}^{(2)}$, $\mathcal{F}_{gg}^{(1)}$, $\mathcal{F}_{gg}^{(2)}$, and $\mathcal{F}_{gg}^{(6)}$. The WW distribution is not directly one of them, but in the Gaussian approximation coupled to the large-$N_c$ limit, which ensures the factorization of CGC expectation values into single trace expectation values, those four gluon distributions can be expressed in terms of $x_2G^{(1)}$ and $x_2G^{(2)}$ \cite{Dominguez:2011wm}: 
\begin{eqnarray}
  {\cal F}_{qg}^{(1)}(x_2,k_t)
  & = & x_2G^{(2)}(x_2,q_t)\,,
  \label{GeneralFqg1}
  \\
  {\cal F}_{qg}^{(2)}(x_2,k_t)
  & = &
  \int d^2q_t\,x_2G^{(1)}(x_2,q_t) F(x_2, k_t-q_t)\,,
  \\
  {\cal F}_{gg}^{(1)}(x_2,k_t)
  & = &
  \int d^2q_t\,x_2G^{(2)}(x_2,q_t) F(x_2, k_t-q_t)\,,
  \\
  {\cal F}_{gg}^{(2)}(x_2,k_t)
  & = &
  -\int d^2q_t\frac{(k_t-q_t)\cdot q_t}{q_t^2}
  x_2G^{(2)}(x_2,q_t) F(x_2,k_t-q_t)\,,
  \\
  {\cal F}_{gg}^{(6)}(x_2,k_t)
  & = &
  \int d^2q_td^2q_t'\,x_2G^{(1)}(x_2,q_t) F(x_2,q_t') F(x_2,k_t-q_t-q_t')\ .
  \label{GeneralFgg6}
\end{eqnarray}
Through (\ref{eq:dipolegluon}) and (\ref{eq:WW(F)}), we have now expressed all the needed gluon TMDs in terms of $F(x_2,k_t)$, the solution of the BK equation in the momentum space (or equivalently Fourier transform of solution of the BK equation in the coordinate space).

\section{Results for the gluon TMDs}

Before we proceed with the computation of the gluon TMDs (\ref{eq:dipolegluon}) and (\ref{GeneralFqg1})-(\ref{GeneralFgg6}) from a solution of the BK equation, we would like to give a couple of useful and interesting results. First, we obtain the gluon TMDs in the Golec-Biernat-Wusthoff (GBW) model analytically; those results may be used for various purposes, such as checking the numerical procedure needed for the various convolution in (\ref{GeneralFqg1})-(\ref{GeneralFgg6}). Second, we compute the high-$k_t$ behavior of the gluon TMDs in the McLerran-Venugopalan (MV), and show that it features the behavior (\ref{eq:highpt}) expected from the operator definitions of the TMDs; this was not obvious a priori, and in general not every model processes this characteristic. We recall that this behavior is necessary in order for the ITMD formula to reproduce the HEF limit when $Q_s\ll k_t\sim P_t$. Finally, we present the gluons with we will use for our cross section calculations, obtained from the KS solution of the BK equation.

\subsection{Analytical results in the GBW model}

The GBW model \cite{GolecBiernat:1998js} is a phenomenological model for the dipole scattering amplitude $N_F(x, {\bf{r}})$, that describes deep inelastic (proton) data at small-$x$ and for moderate values of the photon virtuality. The scattering amplitude in this model is $N_F(x, {\bf{r}}) = 1- \exp\left[-\bold{r}^2 Q_s^2(x)/4\right]$ and the Fourier transform of $S_F$ reads:
\be 
F(x_2,k_t) = \frac{1}{\pi Q_s^2(x_2)}\exp\left[ -\frac{k_t^2}{Q_s^2(x_2)}\right] \ .
\label{eq:F}
\ee
Using this result in formulae (\ref{eq:dipolegluon}) and (\ref{eq:WW(F)}), we get $x_2G^{(1)}$ and $x_2G^{(2)}$ in the GBW model (see appendix~\ref{AppendixGBW}):
\be 
x_2 G^{(2)}(x_2, k_t) = 
\frac{N_c S_\perp}{2\pi^3 \alpha_s Q_s^2(x_2)} k_t^2 \exp\left[ -\frac{k_t^2}{Q_s^2(x_2)}\right]\ ,
\label{G2}
\ee
and
\be 
x_2G^{(1)}(x_2,k_t) = \frac{N_c S_\perp}{4\pi^3 \alpha_s} \int_1^{\infty} \frac{dt}{t}
\exp\left[-\frac{k_t^{2}}{2Q_s^2(x_2)}t\right] \ .
\label{G1Integral}
\ee
Note that the above result for the WW distribution can also be obtained directly from Eq.~\ref{eq:WW_AdjointDipole}, with
$S_A(x, {\bf{r}}) = \exp\left[-\bold{r}^2 Q_s^2(x)/2\right]$. The expression for the WW distribution can be simplified by expressing the remaining integral in
terms of the \emph{exponential integral} special function, $\Ei(x) = \int_x^\infty {dt}\, e^{-t}/t$:
\be 
x_2G^{(1)}(x_2,k_t) = \frac{N_c S_\perp}{4\pi^3 \alpha_s} \Ei\left(\frac{k_t^{2}}{2Q_s^2(x_2)}\right) \ .
\label{G1}
\ee

\begin{figure}[t]
\begin{center}
\includegraphics[width=0.55\textwidth]{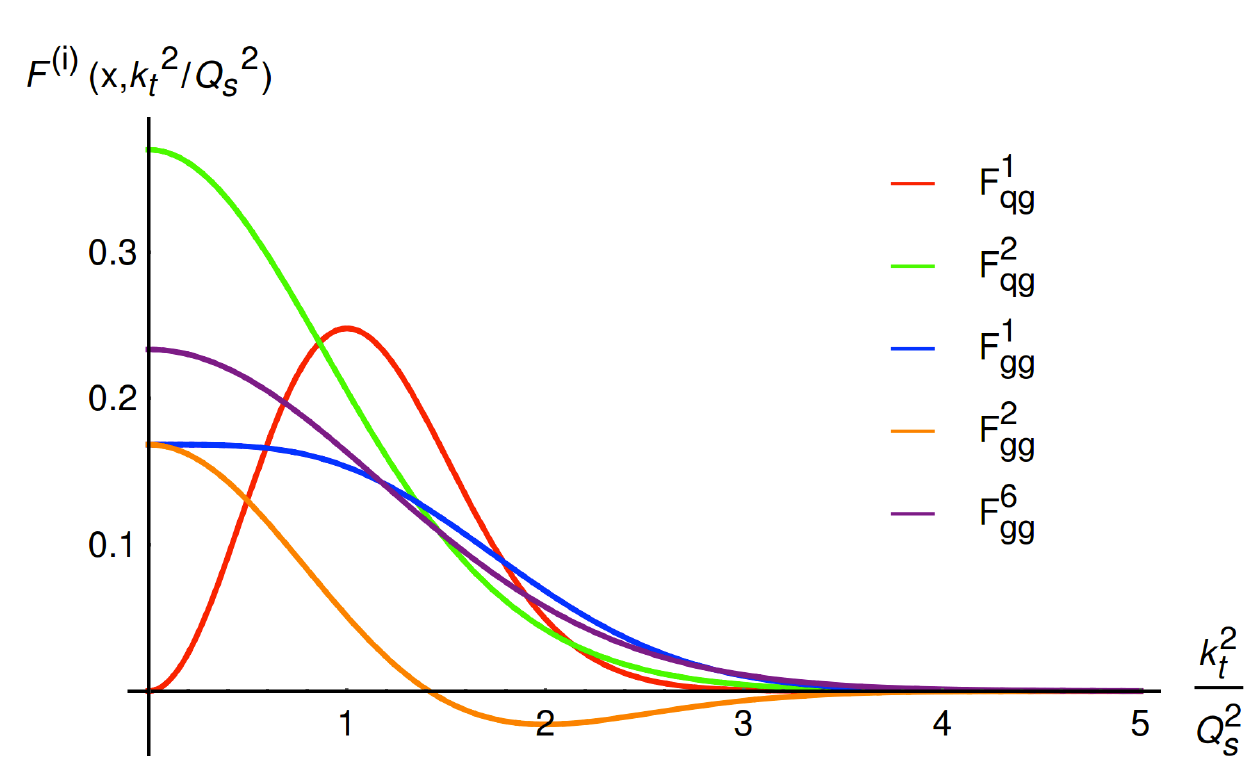}
\hfill
\includegraphics[width=0.38\textwidth]{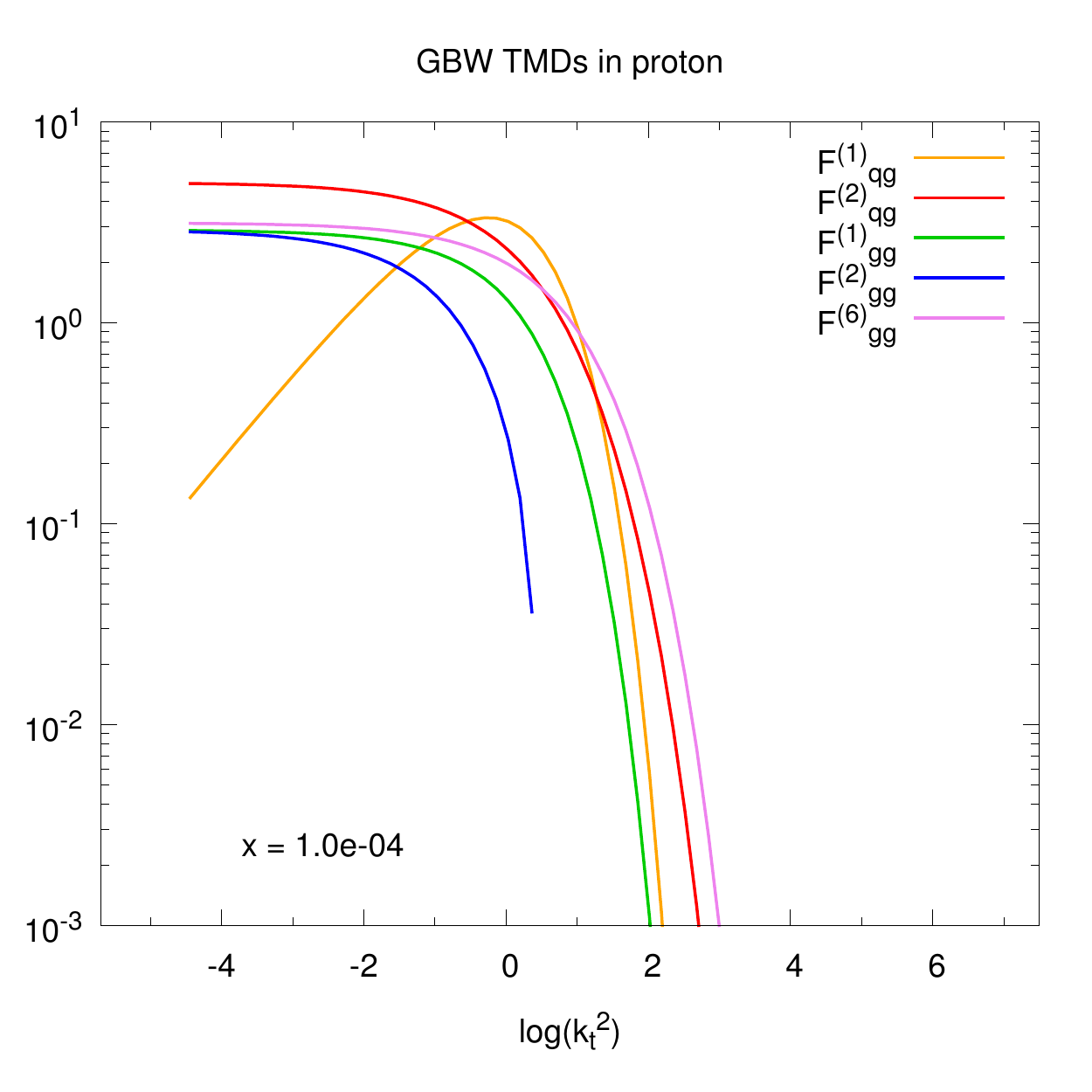}
 \end{center}
\caption{The gluon TMDs (up to a constant factor) in the GBW model as a function of $k_t^2/Q_s^2$ (left) and as a function of $\log (k^2_t/\mbox{GeV}^2)$ at $x=10^{-4}$ (right).}
\label{fig:GBWgluons}
\end{figure}

Using Eqs.~(\ref{eq:F}), (\ref{G2}) and (\ref{G1}) in the relations (\ref{GeneralFqg1})-(\ref{GeneralFgg6}), we get the form of all the gluon TMDs in the GBW model:
\begin{eqnarray}
  \mathcal{F}_{qg}^{(1)}(x_2, k_t)  
  & =  &
  \frac{N_c S_\perp}{2\pi^3 \alpha_s Q_s^2(x_2)} k_t^2 \exp\left[
  -\frac{k_t^2}{Q_s^2(x_2)}\right]\,, \label{FinalFqg1}
  \\
  {\cal F}_{qg}^{(2)}(x_2,k_t)
  & = &
  \frac{N_c S_\perp}{4\pi^3 \alpha_s}
  \left[
    \Ei\left(-\frac{k_t^2}{Q_s^2(x_2)}\right) -
    \Ei\left(-\frac{k_t^2}{3Q_s^2(x_2)}\right)
  \right]\,,
  \\
  {\cal F}_{gg}^{(1)}(x_2,k_t)
  & = &
  \frac{N_c S_\perp}{16\pi^3 \alpha_s}
  \exp\left[-\frac{k_t^{2}}{2Q_s^2(x_2)}\right]\left(2 + \frac{k_t^{2}}{Q_s^2(x_2)} \right)\,,
  \\
  {\cal F}_{gg}^{(2)}(x_2,k_t)
  & = &
  \frac{N_c S_\perp}{16\pi^3 \alpha_s}
  \exp\left[-\frac{k_t^{2}}{2Q_s^2(x_2)}\right]\left(2 - \frac{k_t^{2}}{Q_s^2(x_2)} \right)\,,
  \\
  {\cal F}_{gg}^{(6)}(x_2,k_t) 
  & = &
  \frac{N_c S_\perp}{4\pi^3 \alpha_s}
  \left[
    \Ei\left(-\frac{k_t^2}{2Q_s^2(x_2)}\right) -
    \Ei\left(-\frac{k_t^2}{4Q_s^2(x_2)}\right)
  \right]\,. \label{FinalFgg6}
\end{eqnarray}
Their behavior as a function of $k_t$ is plotted in Fig.~\ref{fig:GBWgluons}, using $Q_s=0.88$ GeV at $x=10^{-4}$.

\subsection{Large-$\mathbf{k_t}$ behavior in the MV model}

The GBW model is not a good parametrization for large transverse momenta, since not only the various TMDs do not converge to a single one at large $k_t$, but also the exponential fall-off of each gluon TMD is unphysical. A model in which those deficiencies are corrected is the MV model
\cite{McLerran:1993ni,McLerran:1993ka}. This model comes about when the color field correlations in the Gaussian approximation are assumed to stay local
$\mu^2(x,{\bold{x}}-{\bold{y}})\to\mu^2\delta({\bold{x}}-{\bold{y}})$. Then the saturation scale is related to the color charge density in the transverse plane of the nucleus $\mu^2$, integrated over the longitudinal direction: $Q_s^2= {g^4C_F}/{(2\pi)}\int dz^+\mu^2$. In addition, the scattering amplitude in this model can be written
\be 
N_F(x, {\bf{r}}) = 1- \exp\left[-\frac{\bold{r}^2 Q_s^2}{4}\log{\frac{1}{\bold r \Lambda}}\right] \ ,
\label{MVdipole}
\ee
where $\Lambda$ is an infrared cut-off.

The logarithmic behavior in (\ref{MVdipole}) is only valid in the limit of small dipole sizes, but this is precisely what we need in order to study the high-$k_t$ behavior of $F(x, {\bf{r}})$ and of the various gluon TMDs. As a matter of fact, the logarithm is the crucial difference between the GBW and the MV model, which restores the correct high-$k_t$ perturbative power-law behavior of the dipole gluon distribution, $x_2 G^{(2)}(x_2, k_t)$, and of the WW gluon distribution, $x_2 G^{(1)}(x_2, k_t)$, which both behave identically as $\sim Q_s^2/k_t^2$ (see for example \cite{Gelis:2001da} and \cite{Kharzeev:2003wz}).

In the appendix \ref{AppendixMV}, we derive the leading order term in $Q_s^2/k_t^2$ for the remaining TMDs,
by expanding Eq.~(\ref{MVdipole}) to first order in $\bold{r}^2 Q_s^2$. We find that, expect for $\mathcal{F}_{gg}^{(2)}$
which goes to zero at leading order, they all scale the same as $x_2 G^{(2)}$ and $x_2 G^{(1)}$:
\bea
\mathcal{F}_{qg}^{(1)}\ , \mathcal{F}_{qg}^{(2)}\ , \mathcal{F}_{gg}^{(1)}\ , \mathcal{F}_{gg}^{(6)} &\simeq& 
\frac{N_c S_\perp Q_s^2}{4\pi^3 \alpha_s k_t^2} + \mathcal{O}\left(\frac{Q_s^4}{k_t^4}\log{\frac{k_t^2}{\Lambda^2}} \right) \ ,
\label{eq:highktbehavior}\\
\mathcal{F}_{gg}^{(2)} &\simeq& 
\mathcal{O}\left(\frac{Q_s^4}{k_t^4}\log{\frac{k_t^2}{\Lambda^2}} \right) \ .
\eea
The sub-leading $N_c$ contribution to $\Phi_{gg\to q\bar{q}}^{(2)}$ (see (\ref{eq:phigg2})) is actually $x_2G^{(1)}$ \cite{Kotko:2015ura}, therefore
these results show that in the MV model, the behavior (\ref{eq:highpt}) is satisfied, and the ITMD formula will indeed reproduce
the HEF limit when $Q_s\ll k_t\sim P_t$. This is also true if the MV model is used as an initial condition to solve the BK equation, since the power-law fall-off (\ref{eq:highktbehavior}) will acquire an anomalous dimension due to the small-$x_2$ evolution, but it will stay the same for all the gluon TMDs.

\subsection{Gluon TMDs from the KS solution to the BK equation}

\begin{figure}[t]
\begin{center}
\includegraphics[width=0.47\textwidth]{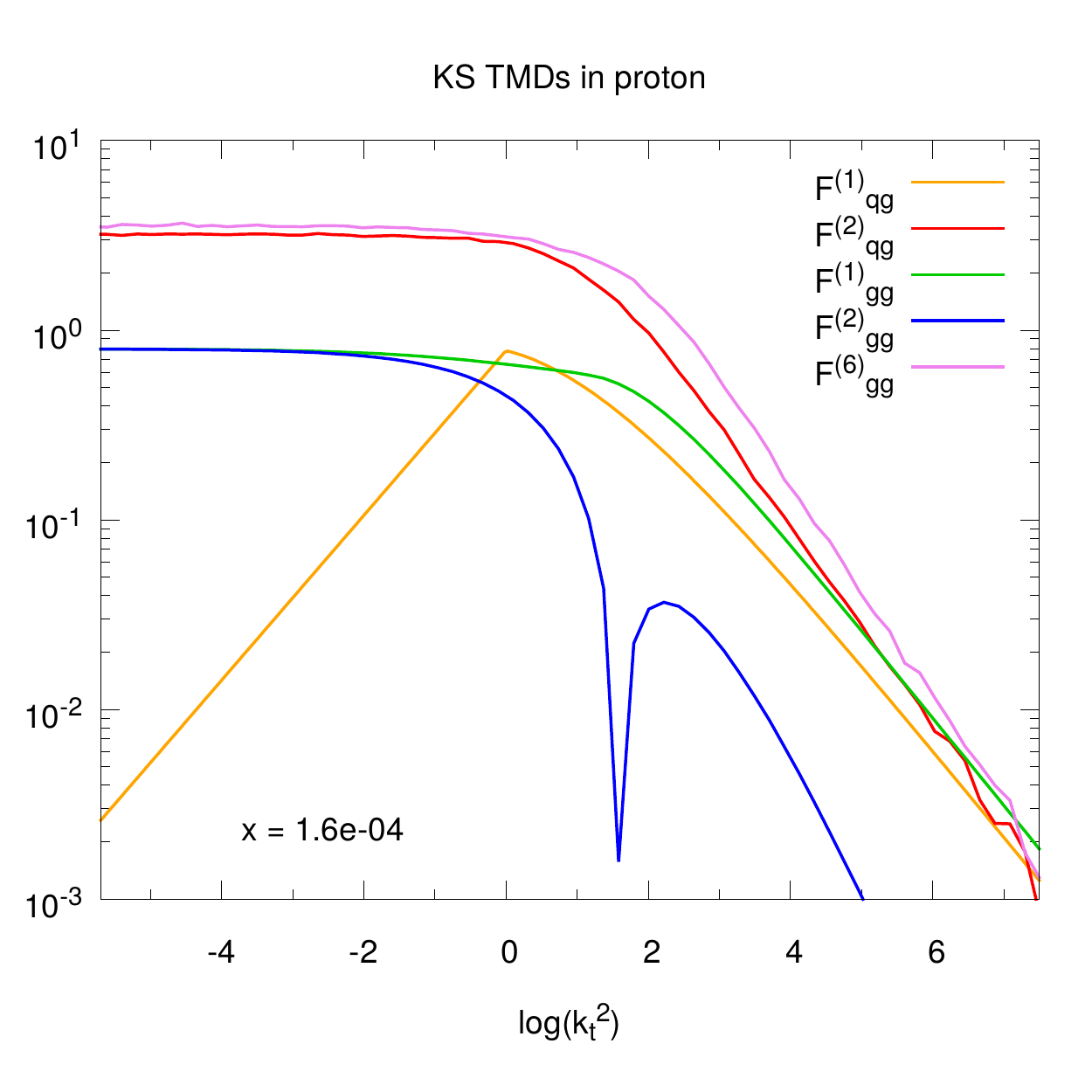}
\hfill
\includegraphics[width=0.47\textwidth]{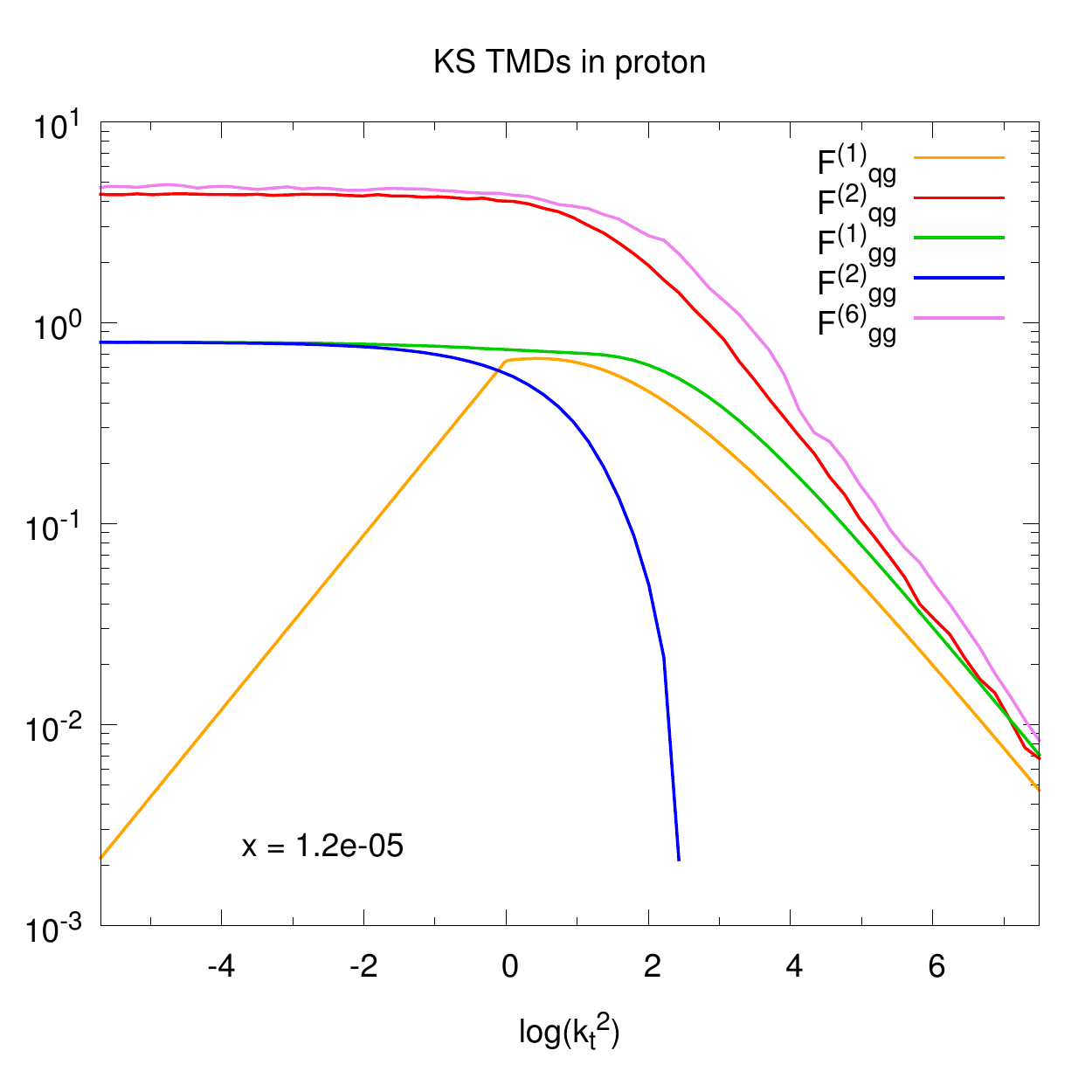}
 \end{center}
\caption{The KS gluon TMDs as a function of $\log (k^2_t/\mbox{GeV}^2)$ at $x=1.6\ 10^{-4}$ for the proton (left) and the lead nucleus (right).
Since ${\cal F}_{gg}^{(2)}$ goes negative, its absolute value is shown on the figures.}
\label{fig:KSgluons}
\end{figure}

The KS solution \cite{Kutak:2012rf} is a solution to BK equation extended to take into account higher-order corrections
relevant in order to provide realistic phenomenological predictions when somewhat large transverse momenta are involved as is the case with jets.
Namely, these are corrections coming from including non-singular pieces of the gluon splitting function, kinematic constraint effects and contributions
from sea quarks \cite{Kutak:2003bd,Kutak:2004ym}. Let us already point out that the initial condition used in \cite{Kutak:2012rf} is not the MV model,
and therefore in the large $k_t$ limit, all the gluon TMDs will not exactly coincide. The mismatch is small however, probably due to the fact that the
initial condition used does also effectively contain a logarithmic behavior.

The KS solution provides directly the dipole gluon TMD $x_2G^{(2)}(x_2,k_t)$, and the parameters of the initial condition are
constrained by a fit to experimental data on deep inelastic scattering off protons. To deal with the nuclear case, the following
formal substitution is made in the non-linear term of the equation
\begin{equation}
\frac{1}{R^2} \to  c\, \frac{A}{R_A^2}\ ,
\qquad {\rm with} \quad
R^2_{\text{A}}= R^2\, A^{2/3}\ ,
\label{eq:radius}
\end{equation}
where $R_A$ is the nuclear radius and $A$ the mass number ($A=208$ for Pb). $c$ is a parameter that we
shall vary between 0.5 and 0.75 in order to assess the uncertainty related to the strength of saturation effect in the
lead nucleus compared to the proton. The nuclear dipole gluon TMD obtained in this way is also normalized to the
number of nucleons $A$. 

In order to calculate all the gluon TMDs (\ref{GeneralFqg1})-(\ref{GeneralFgg6}) from $x_2G^{(2)}(x_2,k_t)$, we are facing the following issue.
The KS solution provides directly an impact-parameter-integrated distribution, which in fact explains why the non-linear term depends on the target size.
As a consequence, it is not straightforward to identify $S_\perp$ and obtain $F(x_2,k_t)$. Our procedure will be to first compute the dipole cross section
$\sigma_{dipole}(x,r\!=\!|{\bf{r}}|)=2\int d^2b\ N_F(x_2, {\bf{r}})$ from $x_2G^{(2)}(x_2,k_t)$ by inverse Fourier transformation of Eq.~(\ref{eq:dipgluontmd}), and then to define $S_\perp$ as its value at large $r$ i.e. when it saturates (since in that limit $N_F\to1$):
\begin{equation}
\frac{1}{2} \sigma_\text{dipole}(x_2,r\!=\!\infty)= S_\perp(x_2)=
\lim_{r \to \infty}
\frac{4\pi^3}{N_c}\alpha_s\int\frac{dk}{k}[1-J_0(k\,r)]\ x_2G^{(2)}(x_2,k)\ .
\end{equation}
We can now obtain $F(x_2,k_t)$ and calculate all the needed gluon TMDs. Their behavior as a function of $k_t$ is plotted in Fig.~\ref{fig:KSgluons}, both for the proton and the lead nucleus. The small mismatch between their high-$k_t$ behavior, expected due to the initial condition for the $x_2$ evolution, can be seen.

%%%%%%%%%%%%%%%%%%%%%%%%%%%%%%%%%%%%%%%%%%%%%%%%%%%%%%%%%%%%%%%%%%%%%%%%%%%% 
\section{Numerical studies of the forward di-jet cross section}
\label{sec:Num_studies}

\begin{figure}[t]
  \begin{center}
    \includegraphics[width=0.48\textwidth]{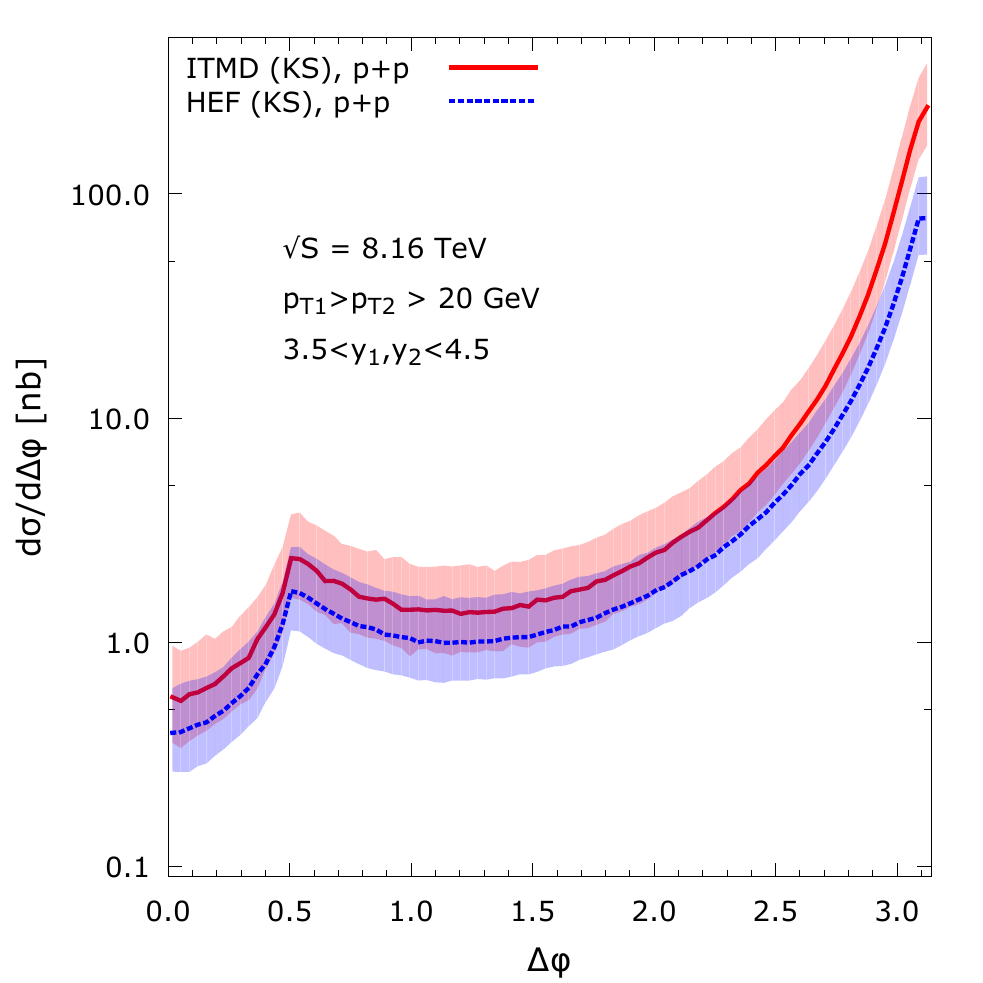}
    \includegraphics[width=0.48\textwidth]{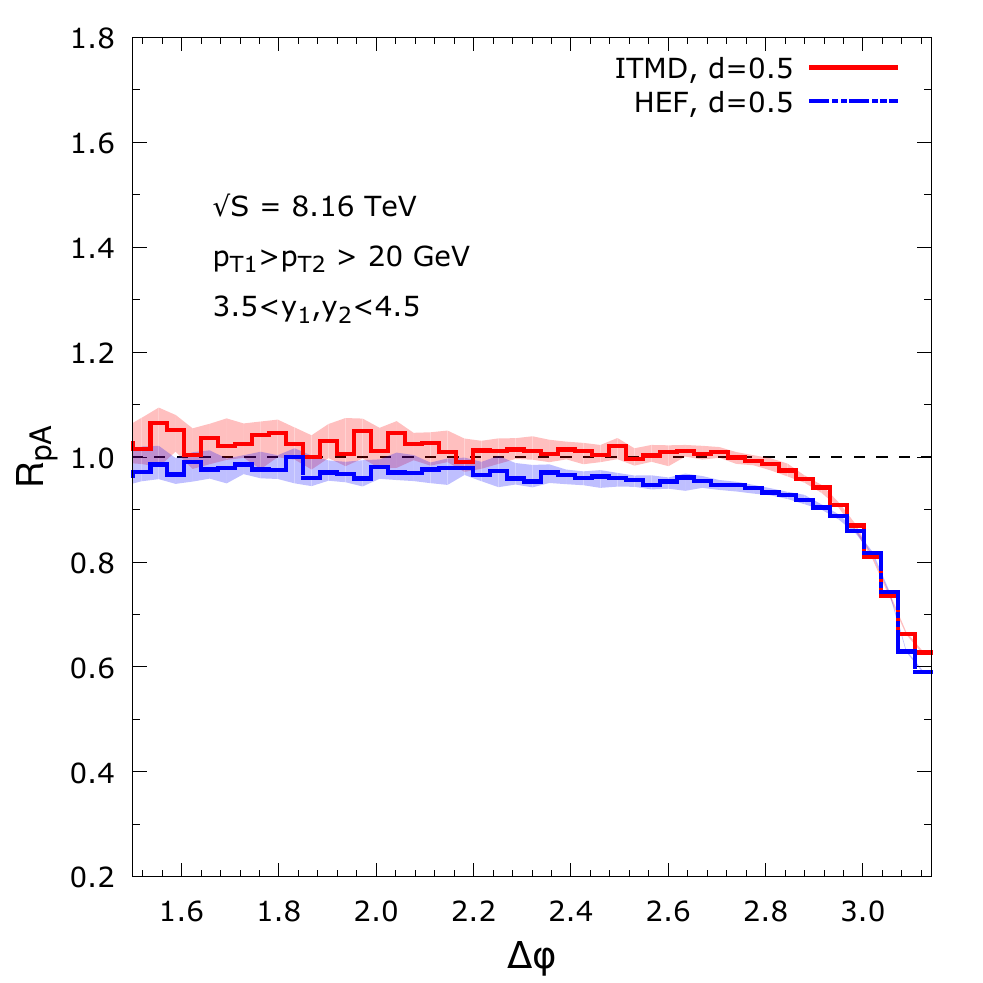}
  \end{center}
  \caption{Left plot: differential cross section as a function of the azimuthal angle between the jets for
  p+p collisions, comparing the new ITMD approach with previously obtained HEF results. The ITMD/HEF difference, which as expected is the largest
  around $\Delta\phi\simeq\pi$ is similar in p+Pb collisions, resulting in almost identical $R_{pPb}$ for both approaches: right plot.}
  \label{fig:dphiHEF-ITMD}
\end{figure}

We move now to the numerical results for forward di-jet production in p+p and p+Pb collisions at the LHC.
We consider a center-of-mass energy of $8.16\, \mathrm{TeV}$, and generate all our predictions with the
forward region defined as the rapidity range $3.5 < y < 4.5$ on one side of the detector. The two hardest
jets are required to lie within this region and we also impose a cut on the minimal transverse momentum
of each two jets: $p_{t0}=20\, \mathrm{GeV}$. In such a setup, the cross section still may be divergent due
to collinear singularities. These are cut-off by applying a jet algorithm on the final state momenta with a
delta-phi-rapidity cut $R=0.5$. Finally, we require the jets to be ordered according to increasing transverse
momentum, that is we have $|p_{t1}|>|p_{t2}|>p_{t0}$.

The new factorization approach summarized in Eq.~\ref{eq:itmd} has been implemented in two independent Monte Carlo
codes \textsc{avhlib} \cite{vanHameren:2007pt,vanHameren:2010gg} and \textsc{LxJet} \cite{LxJet:2013}.
To be more precise, the computer programs do not utilize the formula \ref{eq:itmd} which uses amplitudes squared
and summed over polarizations and colors. Instead, the more generic color-ordered off-shell helicity amplitudes are used,
as derived in \cite{Kotko:2015ura}. This approach follows the modern direction of amplitude calculation and allows for
more 'exclusive' calculations in the future (for example a study of helicity dependence or interfacing with color-flow
dependent parton shower generators). The calculations of matrix elements are made keeping $N_c$ finite.

For the collinear parton distributions that enter the ITMD formula, we chose the general-purpose CT10 set \cite{Lai:2010vv}.
For the central value of the factorization and renormalization scale, we choose the average transverse momentum of the
two leading jets, $\mu_F=\mu_R=\frac12 (|p_{t1}|+|p_{t2}|)$. We will produce error bands corresponding to the renormalization
and factorization scale uncertainties by varying the central numbers from half to twice their value.

For the various observables  ${\cal O}$ shown below, we also consider the nuclear modification factors defined as
\begin{equation}
R_{\rm pPb} = \frac{\displaystyle \frac{d\sigma^{p+Pb}}{d{\cal O}}}
                 {\displaystyle A\ \frac{d\sigma^{p+p}}{d{\cal O}}}\,.
\label{eq:RpA}
\end{equation}
with $A=208$ for Pb. In our approach, in the absence of saturation effects, or in the case in which they are
equally strong in the nucleus and in the proton, this ratio is equal to unity. If,
however, the non-linear evolution plays a more important role in the case of the
nucleus, the $R_{\rm pPb}$ ratio will be suppressed below 1.

\begin{figure}[t]
  \begin{center}
\includegraphics[width=.48\textwidth]{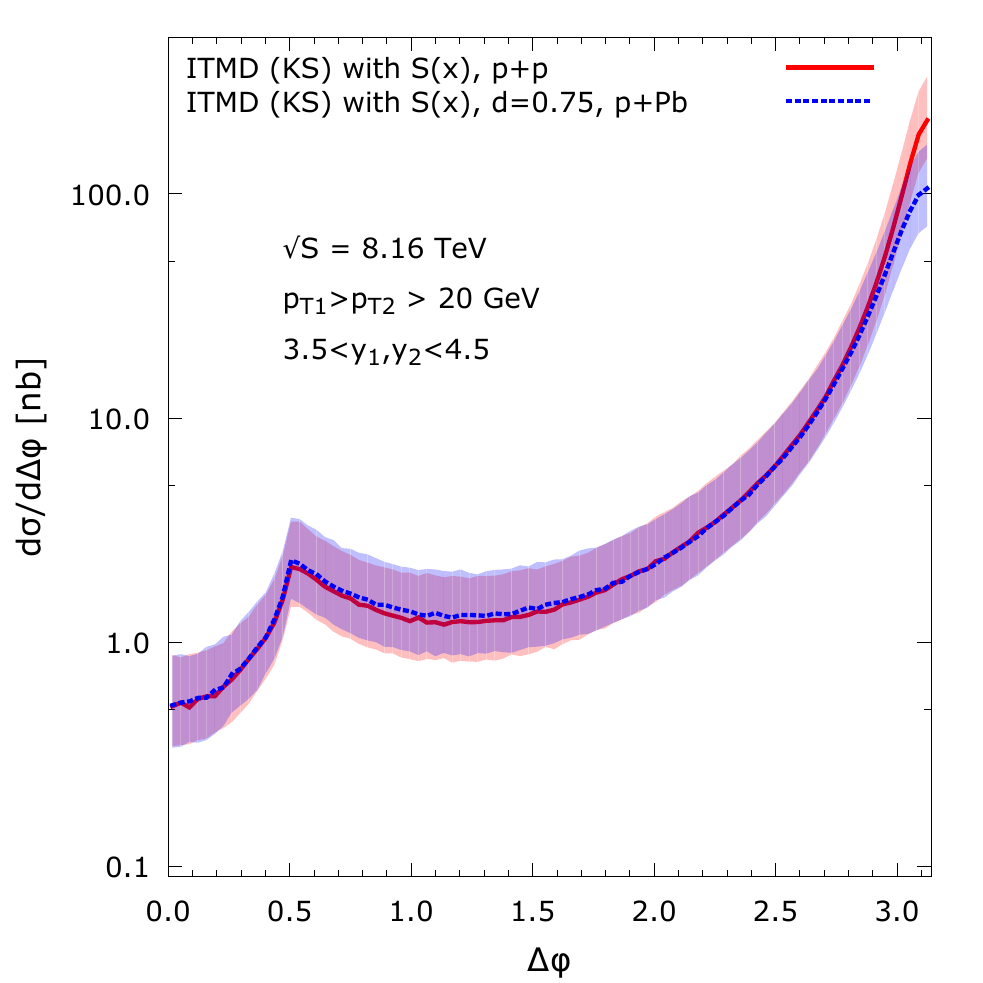}
    \includegraphics[width=0.48\textwidth]{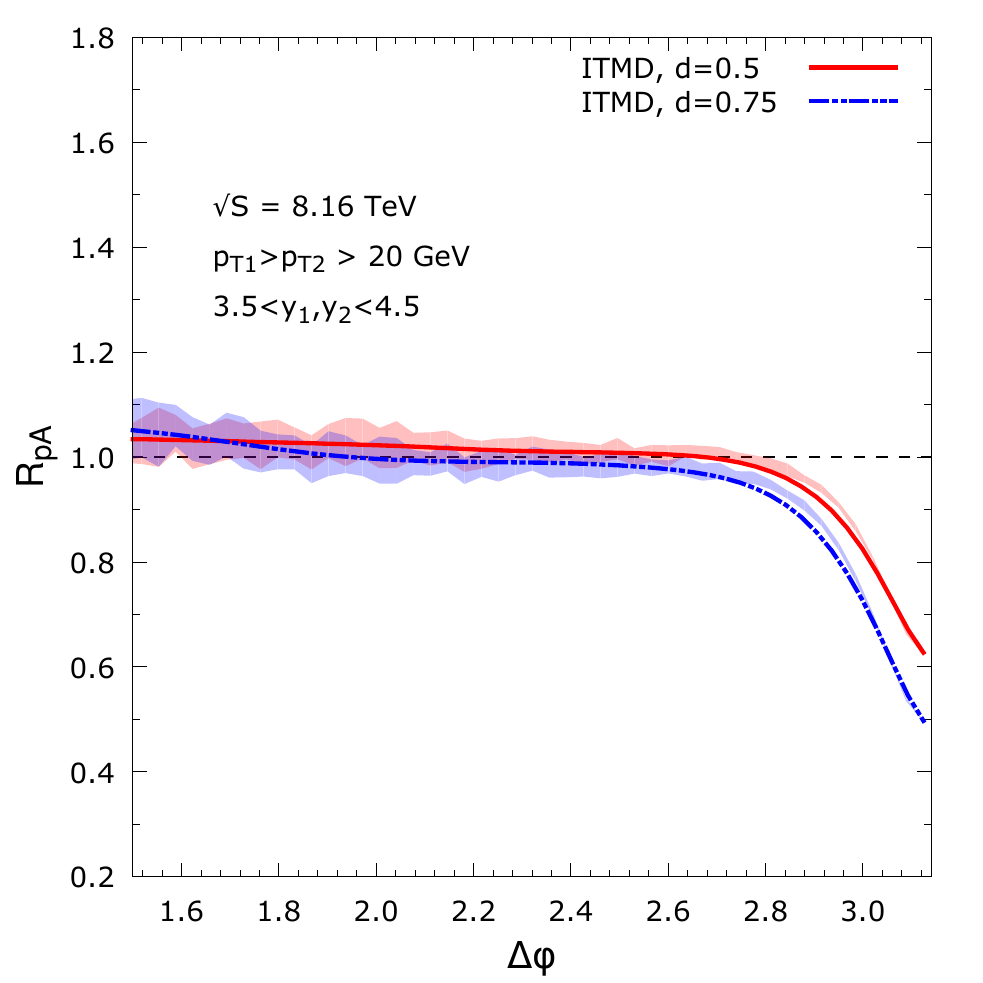}
  \end{center}
  \caption{Left plot: differential cross section as a function of the azimuthal angle between the jets for
  p+p and p+Pb collisions (rescaled by the number of nucleons). The distributions are identical
  everywhere expect near $\Delta\phi\simeq\pi$, where saturation are the strongest. Right plot:
  nuclear modification factors for two values of the nuclear saturation scale, providing an uncertainty band.}
  \label{fig:dphipvsPb}
\end{figure}

We start by investigating the azimuthal correlations, with the azimuthal angle between the jets $\Delta\phi$ defined to lie within $0<\Delta\phi<\pi$.
First we compare the new ITMD approach with previously obtained HEF results in Fig.~\ref{fig:dphiHEF-ITMD}. For the $\Delta\Phi$ distribution in p+p collisions, we see that at small angles where ideally they should match, there remains a small difference between the ITMD and HEF curves. As we anticipated, this is due to the initial condition used to obtain the KS gluons. By contrast, near $\Delta\Phi\simeq\pi$, we observe a large difference, as expected: the ITMD result is about a factor 3 bigger than the HEF one. The ITMD/HEF ratio is very similar in the case of p+Pb collisions, resulting in almost identical $R_{\rm pPb}$ for both approaches, as also shown on the figure. For that comparison, we have parametrized the strength of the non-linear term in the evolution equation for the Pb gluon distributions (see (\ref{eq:radius})) with $c=0.5$.

Next, we compare the $\Delta\Phi$ distribution in p+p and p+Pb collisions in Fig.~\ref{fig:dphipvsPb}. After rescaling the p+Pb cross section by the number of nucleons, we obtain identical distributions almost everywhere. It is only for nearly back-to-back jets, around $\Delta\phi\simeq\pi$, that saturation effects induce a difference. This difference is better appreciated on the nuclear modification factor, which goes from unity to 0.6, as $\Delta\phi$ varies from $\sim2.7$ to $\pi$. Two values of the parameter $c$ have been considered, which makes up an uncertainty band that turns out to be rather small. This means that the uncertainty related to the value of the saturation scale of the lead nucleus does not strongly influence the predicted $R_{\rm pPb}$ suppression.

\begin{figure}[t]
  \begin{center}
    \includegraphics[width=0.48\textwidth]{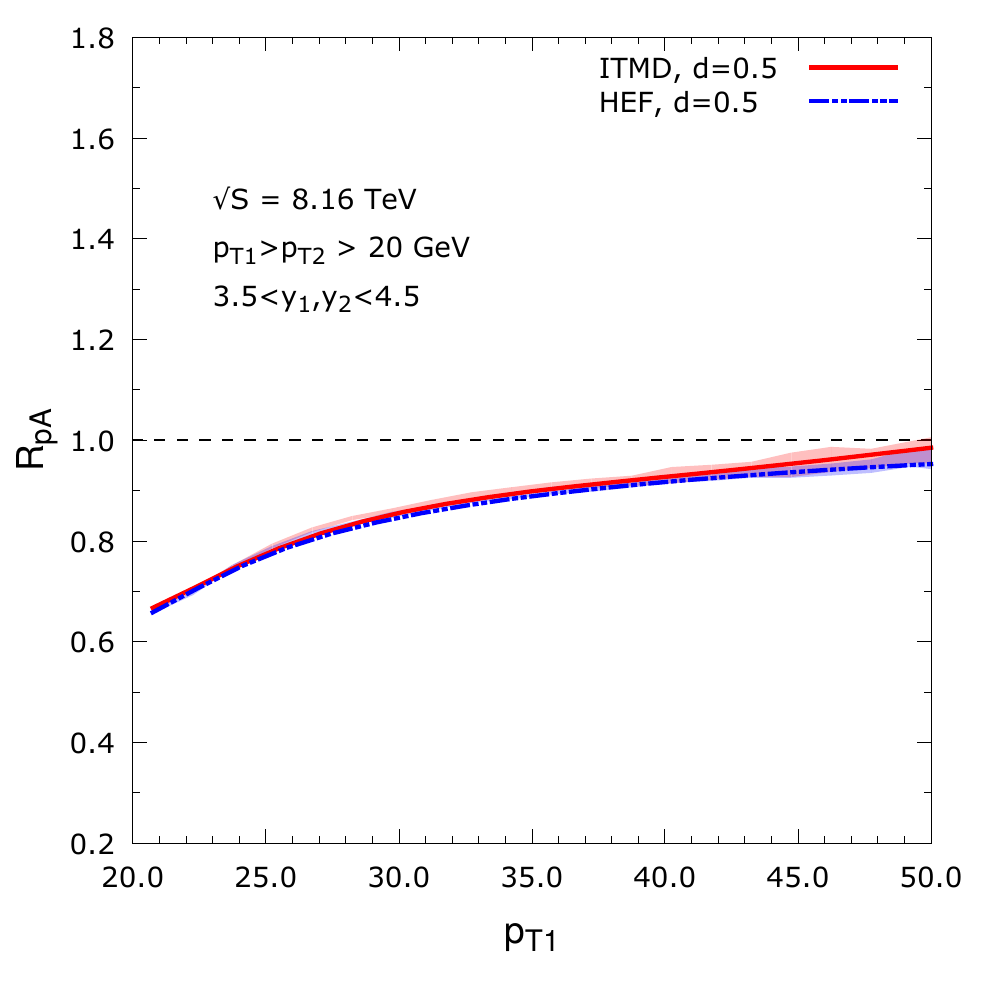}
    \includegraphics[width=0.48\textwidth]{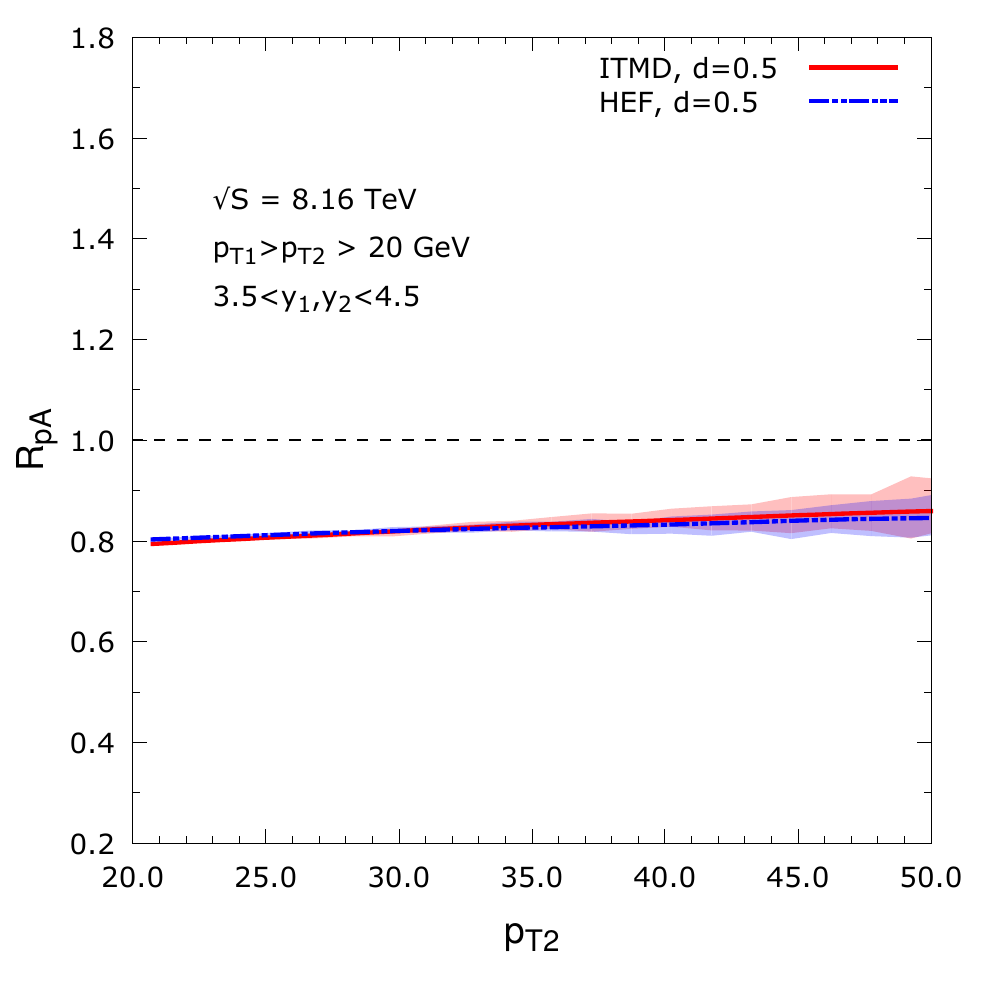}
  \end{center}
  \caption{Nuclear modification factors as a function of the transverse momentum of the leading (left) and subleading (right) jet, 
   comparing the new ITMD approach with previously obtained HEF results.}
  \label{fig:ptHEF-ITMD}
\end{figure}

\begin{figure}[t]
  \begin{center}
    \includegraphics[width=0.48\textwidth]{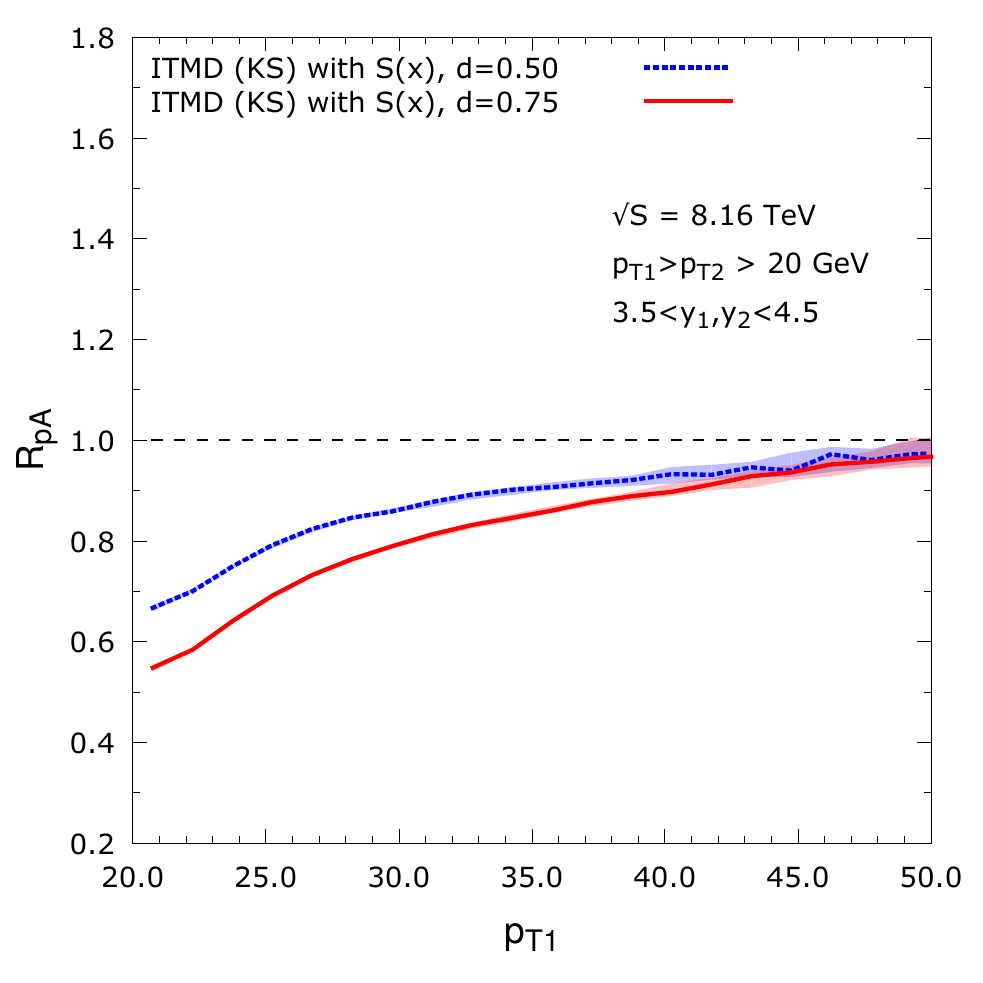}
    \includegraphics[width=0.48\textwidth]{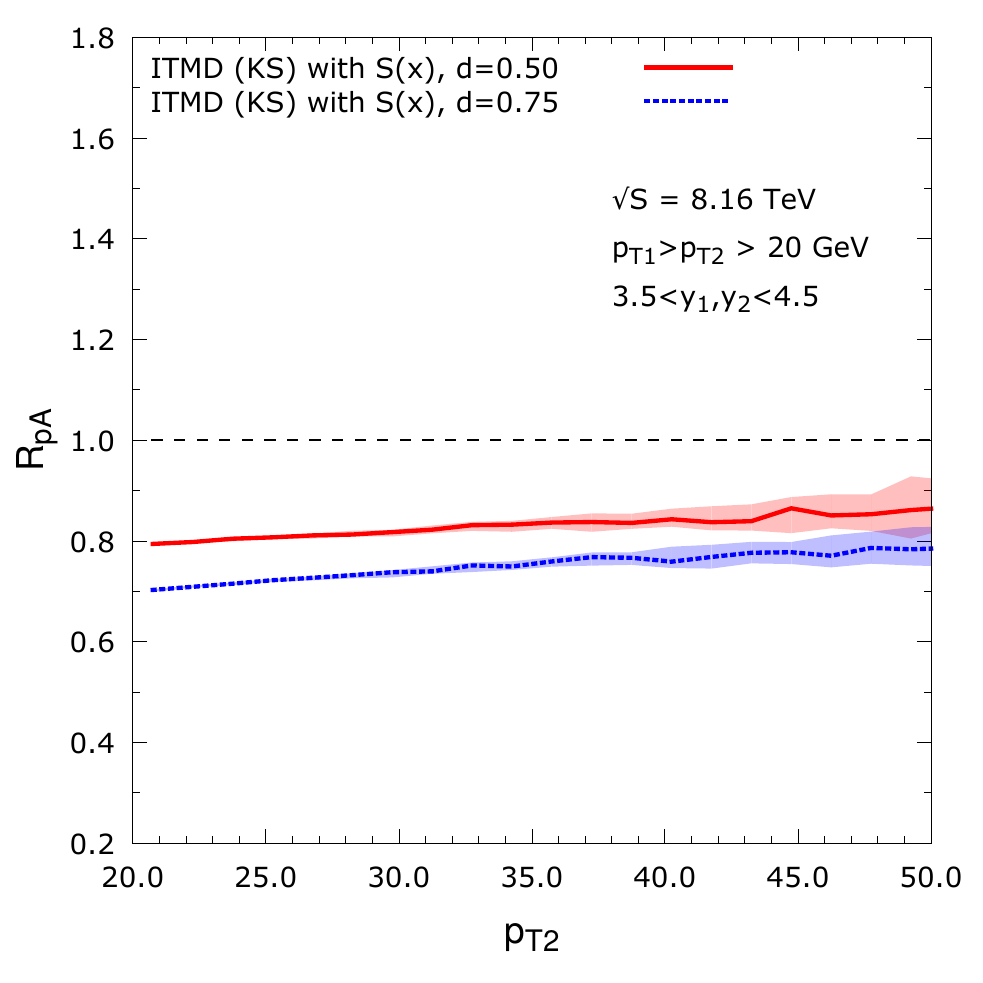}
  \end{center}
  \caption{Comparison of nuclear modification factors as a function of the
  transverse momentum of the leading (left) and subleading (right) jet, using two with different
  choices of the non-linear term strength parameter for the nucleus $c$.}
  \label{fig:ptpvsPb}
\end{figure}

Finally, in Fig.~\ref{fig:ptHEF-ITMD} we display the nuclear modification factors as a function
of the transverse momentum of the leading and sub-leading jet. Our conclusions are similar for these observables: the new ITMD
predictions are similar to the previously obtained HEF results, due to the fact that the ITMD/HEF ratio is similar in p+p and p+Pb collisions.
This means that the HEF framework, which is incorrect for nearly back-to-back jets - since in this formalism all the gluon TMDs are considered
equal regardless of the kinematics - can nevertheless be safely used for $R_{\rm pPb}$ calculations. The same is not true for cross section
calculations. Fig.~\ref{fig:ptpvsPb} shows those same nuclear modification factors but comparing the predicted suppression for two different
values of the parameter $c$. As a function of the leading jet $p_t$, $R_{\rm pPb}$ rises up from about 0.6 for $p_{t1}=20$ GeV to unity for $p_{t1}=50$ GeV.
However, it is interesting to note that as a function of the sub-leading jet $p_t$, this ratio rather stays flat around $0.8$.

\section{Conclusions}
  
In this paper, we have studied forward di-jet production in proton-proton and proton-lead collisions,
using the small-x improved TMD factorization framework Eq.~(\ref{eq:itmd}. We have obtained the first 
numerical implementation of this formalism, and the first predictions for forward di-jets at the LHC, a process
which is particularly interesting from small-$x$ point of view. Our results for the nuclear modification factors
in p+Pb vs p+p collisions confirm the conclusions obtained in \cite{vanHameren:2014lna} in the HEF framework,
that for nearly back-to-back jets, non negligible effects of gluon saturation are to be expected as
one goes from p+p to p+Pb collisions.

This is due to the fact that in such configurations, the total transverse momentum of the jet pair is of the
order of the saturation scale of the target, and even though the jet transverse momenta are individually much larger than
$Q_s$, saturation effects are not irrelevant. To obtain our predictions, we used the KS gluon distributions,
and it would certainly be interesting use other extensions of the BK equation, such as for instance the rcBK gluon
distribution \cite{Albacete:2010sy}, in order to compare the level of saturation effects expected.

It is important to note that so far, our results have been obtained using an impact-parameter
averaged nuclear saturation scale. However, the outcome of high-energy proton-nucleus
collisions seems to be quite sensitive to the fact that the nucleon positions in the nucleus fluctuate
event by event. We have provided predictions using two different nuclear saturation
strength parameter $c$, but a more complete study including such nucleon-level fluctuation effects
would allow to better estimate the uncertainty related to the nuclear geometry. In the meantime, our
results are enough to motivate experimental measurements at the LHC.

Finally, one important theoretical ingredient is still missing in our formulation: the Sudakov logarithms.
Their effect should be the largest also for nearly back-to-back jets, since they are logarithms of the
ratio of the hard scale to the transverse momentum imbalance of the jet pair, and therefore are expected to compete
with saturation effects. The Sudakov logarithms are identical in p+p and p+Pb collisions, which means
that to some extent they will cancel in the nuclear modification factors, as has been observed in an HEF based
approach in \cite{Kutak:2014wga,vanHameren:2014ala}, but nevertheless they could smear the saturation effects,
depending on which contributes the most. We plan to tackle those interesting studies in the future.
  
\section*{Acknowledgements}
K.K acknowledges support by Narodowe Centrum Nauki with Sonata Bis grant DEC-2013/10/E/ST2/00656. The work of P.K. was supported by the Department of Energy Grants No. DE-SC-0002145, DE-FG02-93ER40771. A.v.H.\ was supported by grant of National Science Center, Poland, No.\ 
2015/17/B/ST2/01838.

%%%%%%%%%%%%%%%%%%%%%%%%%%%%%%%%%%%%%%%%%%%%%%%%%%%%%%%%%%%%%%%%%%%%%%%%%%%% 
\appendix

\section{Calculation of the gluon distributions in the GBW model}
\label{AppendixGBW}

In this appendix we list some of the intermediate steps leading to the result (\ref{G1Integral}). The Weizs{\"a}cker-Williams gluon distribution can be obtained from Eq.~\ref{eq:WW(F)} by using Eqs.~(\ref{eq:F}) and~(\ref{G2}):
\bea
x_2G^{(1)}(x_2,k_t) &=& \frac{N_c S_\perp}{4\pi^4 \alpha_s Q_s^4(x_2)}
\int_{k_t^2}^{\infty} dk_t^{'2}\ln\left(\frac{k_t'^{2}}{k_t^2}\right)
\int d^2 q_t \exp\left[ -\frac{q_t^2}{Q_s^2(x_2)}\right] \exp\left[ -\frac{(k_t^{'} - q_t)^2}{Q_s^2(x_2)}\right] \nonumber \\
&=&
%\frac{N_c S_\perp}{4\pi^4 \alpha_s Q_s^4(x_2)}
 %\int_{k_t^2}^{\infty} dk_t^{'2}\ln\left(\frac{k_t'^{2}}{k_t^2}\right)
%\exp\left[ -\frac{k_t^{'2}}{Q_s^2(x_2)}\right]   \nonumber \\
%&\times &
%\int_0^{\infty} d |q_t| \, |q_t| \exp\left[ -2\frac{q_t^2}{Q_s^2(x_2)}\right]
%\int_0^{2\pi} d\phi \exp\left[ 2\frac{|k_t^{'}||q_t|\cos{\phi}}{Q_s^2(x_2)}\right] \nonumber \\
%&=&
\frac{N_c S_\perp}{2\pi^3 \alpha_s Q_s^4(x_2)}
 \int_{k_t^2}^{\infty} dk_t^{'2}\ln\left(\frac{k_t'^{2}}{k_t^2}\right)
\exp\left[ -\frac{k_t^{'2}}{Q_s^2(x_2)}\right]   \nonumber \\
&\times &
\int_0^{\infty} d |q_t| \, |q_t| \exp\left[ -2\frac{q_t^2}{Q_s^2(x_2)}\right] I_0 \left(2\frac{|k_t^{'}||q_t|}{Q_s^2(x_2)}\right) \nonumber \\
&=&
\frac{N_c S_\perp}{8\pi^3 \alpha_s Q_s^2(x_2)}
 \int_{k_t^2}^{\infty} dk_t^{'2}\log\left(\frac{k_t'^{2}}{k_t^2}\right)
 \exp\left[ -\frac{k_t^{'2}}{2Q_s^2(x_2)}\right]\ .
\eea
By performing a change of variables $k_t' \to k_t t$ we get: 
\be 
x_2G^{(1)}(x_2,k_t) = \frac{N_c S_\perp}{8\pi^3 \alpha_s Q_s^2(x_2)} k_t^2 \int_1^{\infty} dt \log(t)
\exp\left[-\frac{k_t^{2}}{2Q_s^2(x_2)}t\right] \ .
\ee
One partial integration leads to the final result for $x_2G_1(x_2,k_t)$:
\be 
x_2G^{(1)}(x_2,k_t) = \frac{N_c S_\perp}{4\pi^3 \alpha_s} \int_1^{\infty} \frac{dt}{t}
\exp\left[-\frac{k_t^{2}}{2Q_s^2(x_2)}t\right]\ .
\ee
Similar calculations give the results (\ref{FinalFqg1})-(\ref{FinalFgg6}).

\section{High-$k_t$ limit of the gluon distributions in the MV model}
\label{AppendixMV}

The large-$k_t$ behavior of the dipole and WW distributions has been derived before (see for instance \cite{Gelis:2001da} and \cite{Kharzeev:2003wz}). We briefly recall these results and then we calculate ${\cal F}_{qg}^{(i)}$ and ${\cal F}_{gg}^{(i)}$. The basic building block is the Fourier transform of the fundamental dipole
\be 
F(x_2,k_t) = \int \frac{d^2{\bf{r}}}{(2\pi)^2} e^{-i{k_t}\cdot{\bf r}} S_F(x, {\bf{r}}) \ .
\ee
We are interested in the region of large transverse momentum, or equivalently in small values of dipole sizes ${\bf r}$. Therefore, we expand $S_F$ to first non-trivial order in ${\bf r}^2 Q_s^2$:
\be 
F(x_2,k_t) \simeq \int \frac{d^2{\bf{r}}}{(2\pi)^2} e^{-i{k_t}\cdot{\bf r}} \left[ 1 - \frac{r^2 Q_s^2(x_2)}{4}\log{\frac{1}{r \Lambda}} + 
\mathcal{O}\left(r^4 Q_s^4(x_2)\log^2{\frac{1}{ r \Lambda}}\right)\right] \ .
\ee
The first term in the expansion formally gives a delta function $\delta^{(2)}(k_t)$. This term will contribute only for values of $k_t$ around zero, and not in the region of large $k_t$ that is considered here, so it can be safely dropped. The next terms give:
\bea 
F(x_2,k_t) &\simeq & -\frac{ Q_s^2(x_2)}{4(2\pi)^2}\int d^2{\bf{r}} e^{-i{k_t}\cdot{\bf r}}\, r^2\log{\frac{1}{ r \Lambda}} \left[1 + 
\mathcal{O}\left(r^2 Q_s^2(x_2)\log{\frac{1}{ r \Lambda}}\right)\right] \nonumber \\
&\simeq &
-\frac{ Q_s^2(x_2)}{8\pi}\int dr \, r^3 \log{\frac{1}{r \Lambda}}\left[1 + 
\mathcal{O}\left(r^2 Q_s^2(x_2)\log{\frac{1}{ r \Lambda}}\right)\right]J_0(k_t r) \nonumber \\
&\simeq &
\frac{1}{2\pi}\frac{Q_s^2(x_2)}{k_t^4} + \mathcal{O}\left(\frac{Q_s^4(x_2)}{k_t^6}\log{\frac{k_t^2}{\Lambda^2}} \right) \ .
\eea
The dipole gluon distribution from Eq.~(\ref{eq:dipolegluon}) is:
\be 
{\cal F}_{qg}^{(1)}(x_2,k_t) \equiv x_2G^{(2)}(x_2,k_t) \simeq 
\frac{N_c S_\perp Q_s^2(x_2)}{4\pi^3 \alpha_s k_t^2} + \mathcal{O}\left(\frac{Q_s^4(x_2)}{k_t^4}\log{\frac{k_t^2}{\Lambda^2}} \right) \ .
\ee 

Similarly, to get the perturbative behavior of the WW density, we start from Eq.~(\ref{eq:WW_AdjointDipole}), and we expand the adjoint dipole,
$S_A(x,r)\!\simeq\!1\!-\!\frac{r^2 Q_s^2}{2}\log{\frac{1}{r \Lambda}}\!+\!\mathcal{O}\left(r^4 Q_s^4\log^2{\frac{1}{ r \Lambda}}\right)$. For $x_2G^{(1)}$ we get:
\bea
x_2G^{(1)}(x_2,k_t) &=&  \frac{N_c S_\perp}{4 \alpha_s \pi^4}\frac{Q_s^2(x_2)}{2}\int d^2{\bf r}\
e^{-i{k_t}\cdot{\bf r}}\ \left[\log{\frac{1}{r \Lambda}} + 
\mathcal{O}\left(r^2 Q_s^2(x_2)\log^2{\frac{1}{ r \Lambda}}\right)\right] \nonumber \\
&=&
\frac{N_c S_\perp}{2 \alpha_s \pi^3}\frac{Q_s^2(x_2)}{2}\int d r \, r \ \left[\log{\frac{1}{r \Lambda}} + 
\mathcal{O}\left(r^2 Q_s^2(x_2)\log^2{\frac{1}{ r \Lambda}}\right)\right] J_0(k_tr) \nonumber \\
&=&
\frac{N_c S_\perp Q_s^2(x_2)}{4\pi^3 \alpha_s k_t^2} + \mathcal{O}\left(\frac{Q_s^4(x_2)}{k_t^4}\log{\frac{k_t^2}{\Lambda^2}} \right)\ .
\eea

Using the above results we can calculate the perturbative expansion of the rest of the distributions. For $\mathcal{F}_{qg}^{(2)}$ we have:
\bea
\mathcal{F}_{qg}^{(2)}(x_2, k_t) &=&
\int d^2q_t\,x_2G^{(1)}(x_2,q_t) F(x_2, k_t-q_t) \nonumber \\
&\simeq&
\int d^2q_t\, \left[ \frac{N_c S_\perp Q_s^2(x_2)}{4\pi^3 \alpha_s q_t^2} + \mathcal{O}\left(\frac{Q_s^4(x_2)}{q_t^4}\log{\frac{q_t^2}{\Lambda^2}} \right) \right] \nonumber \\
&\times &
\left[\delta^{(2)}(k_t - q_t) + \frac{1}{2\pi}\frac{Q_s^2(x_2)}{(k_t-q_t)^4} + \mathcal{O}\left(\frac{Q_s^4(x_2)}{(k_t-q_t)^6}\log{\frac{(k_t-q_t)^2}{\Lambda^2}} \right)\right] \nonumber \\
&\simeq & 
\frac{N_c S_\perp Q_s^2(x_2)}{4\pi^3 \alpha_s k_t^2} + \mathcal{O}\left(\frac{Q_s^4(x_2)}{k_t^4}\log{\frac{k_t^2}{\Lambda^2}} \right)\ .
\eea
To this order $x_2G^{(1)}(x_2, k_t) = x_2G^{(2)}(x_2, k_t)$ and
\be 
\mathcal{F}_{gg}^{(1)}(x_2, k_t) = \mathcal{F}_{qg}^{(2)}(x_2, k_t) = \frac{N_c S_\perp Q_s^2(x_2)}{4\pi^3 \alpha_s k_t^2} + \mathcal{O}\left(\frac{Q_s^4(x_2)}{k_t^4}\log{\frac{k_t^2}{\Lambda^2}} \right)\ .
\ee
Similarly:
\bea 
\mathcal{F}_{gg}^{(2)}(x_2, k_t) &=& 
-\int d^2q_t\frac{(k_t-q_t)\cdot q_t}{q_t^2}
  x_2G^{(2)}(x_2,q_t) F(x_2,k_t-q_t) \nonumber \\
&\simeq &
-\int d^2q_t\frac{(k_t-q_t)\cdot q_t}{q_t^2}   \left[\frac{N_c S_\perp Q_s^2(x_2)}{4\pi^3 \alpha_s q_t^2} + \mathcal{O}\left(\frac{Q_s^4(x_2)}{q_t^4}\log{\frac{q_t^2}{\Lambda^2}} \right)\right] \nonumber \\
&\times &
\left[\delta^{(2)}(k_t - q_t) + \frac{1}{2\pi}\frac{Q_s^2(x_2)}{(k_t-q_t)^4} + \mathcal{O}\left(\frac{Q_s^4(x_2)}{(k_t-q_t)^6}\log{\frac{(k_t-q_t)^2}{\Lambda^2}} \right)\right] \nonumber \\
&\simeq & 0 \ ,
\eea
and
\bea
{\cal F}_{gg}^{(6)}(x_2,k_t)
  &=&
  \int d^2q_td^2q_t'\,x_2G^{(1)}(x_2,q_t) F(x_2,q_t') F(x_2,k_t-q_t-q_t') \nonumber \\
& \simeq &
\int d^2q_td^2q_t'\,  
\left[ \frac{N_c S_\perp Q_s^2(x_2)}{4\pi^3 \alpha_s q_t^2} + \mathcal{O}\left(\frac{Q_s^4(x_2)}{q_t^4}\log{\frac{q_t^2}{\Lambda^2}} \right) \right] \nonumber \\
&\times &
\left[\delta^{(2)}(q_t') + \frac{1}{2\pi}\frac{Q_s^2(x_2)}{q_t'^4} + \mathcal{O}\left(\frac{Q_s^4(x_2)}{q_t'^6}\log{\frac{q_t'^2}{\Lambda^2}} \right)\right] \nonumber \\
& \times &
\left[\delta^{(2)}(k_t - q_t-q_t') + \frac{1}{2\pi}\frac{Q_s^2(x_2)}{(k_t-q_t-q_t')^4} +
\mathcal{O}\left(\frac{Q_s^4(x_2)}{(k_t-q_t-q_t')^6}\log{\frac{(k_t-q_t-q_t')^2}{\Lambda^2}} \right)\right]\nonumber \\
& \simeq &
\frac{N_c S_\perp Q_s^2(x_2)}{4\pi^3 \alpha_s k_t^2} + \mathcal{O}\left(\frac{Q_s^4(x_2)}{k_t^4}\log{\frac{k_t^2}{\Lambda^2}} \right)\ .
\eea

%%%%%%%%%%%%%%%%%%%%%%%%%%%%%%%%%%%%%%%%%%%%%%%%%%%%%%%%%%%%%%%%%%%%%%%%%%%% 

\end{document}